\PassOptionsToPackage{unicode}{hyperref}
\PassOptionsToPackage{hyphens}{url}
\documentclass[
  fleqn]{article}
\usepackage{lmodern}
\usepackage{amssymb,amsmath}
\usepackage{ifxetex,ifluatex}
\ifnum 0\ifxetex 1\fi\ifluatex 1\fi=0 
  \usepackage[T1]{fontenc}
  \usepackage[utf8]{inputenc}
  \usepackage{textcomp} 
\else 
  \usepackage{unicode-math}
  \defaultfontfeatures{Scale=MatchLowercase}
  \defaultfontfeatures[\rmfamily]{Ligatures=TeX,Scale=1}
\fi
\IfFileExists{upquote.sty}{\usepackage{upquote}}{}
\IfFileExists{microtype.sty}{
  \usepackage[]{microtype}
  \UseMicrotypeSet[protrusion]{basicmath} 
}{}
\makeatletter
\@ifundefined{KOMAClassName}{
  \IfFileExists{parskip.sty}{%
    \usepackage{parskip}
  }{
    \setlength{\parindent}{0pt}
    \setlength{\parskip}{6pt plus 2pt minus 1pt}}
}{
  \KOMAoptions{parskip=half}}
\makeatother
\usepackage{xcolor}
\IfFileExists{xurl.sty}{\usepackage{xurl}}{} 
\IfFileExists{bookmark.sty}{\usepackage{bookmark}}{\usepackage{hyperref}}
\hypersetup{
  pdftitle={Efficient Portfolios},
  pdfauthor={Keith A. Lewis},
  hidelinks,
  pdfcreator={LaTeX via pandoc}}
\usepackage{amsmath,bm}

\newlength{\cslhangindent}
\newenvironment{cslreferences}%
  {\setlength{\parindent}{0pt}%
  \everypar{\setlength{\hangindent}{\cslhangindent}}\ignorespaces}%
  {\par}

\title{Efficient Portfolios\thanks{Peter Carr and David Shimko gave
insightful feedback to make the exposition more accessible to finance
professionals. Any remaining infelicities or omissions are my fault.}}
\author{Keith A. Lewis}
\date{September 22, 2020}

\begin{document}
\maketitle

Given two random realized returns on an investment, which is to be
preferred? This is a fundamental problem in finance that has no
definitive solution except in the case one investment always returns
more than the other. In 1952 Markowitz(Markowitz
\protect\hyperlink{ref-Mar52}{1952}) and Roy(Roy
\protect\hyperlink{ref-Roy52}{1952}) introduced the following criterion
for risk vs.~return in portfolio selection: if two portfolios have the
same expected realized return then prefer the one with smaller variance.
An \emph{efficient portfolio} has the least variance among all
portfolios having the same expected realized return.

In the one-period model every efficient portfolio belongs to a
two-dimensional subspace of the set of all possible realized returns and
is uniquely determined given its expected realized return. We show that
if \(R\) is the (random) realized return of any efficient portfolio and
\(R_0\) and \(R_1\) are the realized returns of any two linearly
independent efficient portfolios then \[
    R - R_0 = \beta(R_1 - R_0)
\] where
\(\beta = \operatorname{Cov}(R - R_0, R_1 - R_0)/\operatorname{Var}(R_1 - R_0)\).
This generalizes the classical Capital Asset Pricing Model formula for
the expected realized return of efficient portfolios. Taking expected
values of both sides when \(\operatorname{Var}(R_0) = 0\) and \(R_1\) is
the ``market'' portfolio gives \[
    E[R] - R_0 = \beta(E[R_1] - R_0)
\] where \(\beta = \operatorname{Cov}(R, R_1)/\operatorname{Var}(R_1)\).

The primary contribution of this short note is observation that the CAPM
formula holds for realized returns as random variables, not just their
expectations. This follows directly from writing down a mathematical
model for one period investments.

\hypertarget{one-period-model}{%
\subsection{One-Period Model}\label{one-period-model}}

The one-period model is parameterized directly by instrument prices.
These have a clear financial interpretation and all other relevant
financial quantities can be defined in terms of prices and portfolios.

Let \(I\) be the set of \emph{market instruments} and \(\Omega\) be the
set of possible market outcomes over a single period. The
\emph{one-period model} specifies the initial instrument \emph{prices}
\(x\in\bm{R}^I\) and the final instrument prices
\(X\colon\Omega\to\bm{R}^I\) depending on the outcome. We assume, as
customary, that there are no cash flows associated with the instruments
and transactions are perfectly liquid and divisible. The one period
model also specifies a probability measure \(P\) on the space of
outcomes.

It is common in the literature to write \(\bm{R}^n\) instead of
\(\bm{R}^I\) where \(n\) is the cardinality of the set of instruments
\(I\). If \(A^B = \{f\colon B\to A\}\) is the set of functions from
\(B\) to \(A\) then \(x\in\bm{R}^I\) is a function
\(x\colon I\to\bm{R}\) where \(x(i) = x_i \in\bm{R}\) is the price of
instrument \(i\in I\). This avoids circumlocutions such as let
\(I = \{i_1,\ldots,i_n\}\) be the set of instruments and
\(x = (x_1,\ldots,x_n)\) be their corresponding prices \(x_j = x(i_j)\),
\(j = 1,\ldots, n\).

A \emph{portfolio} \(\xi\in\bm{R}^I\) is the number of shares initially
purchased in each instrument. The \emph{value} of a portfolio \(\xi\)
given prices \(x\) is \(\xi\cdot x = \sum_{i\in I}\xi_i x_i\). It is the
cost of attaining the portfolio \(\xi\). The \emph{realized return} is
\(R(\xi) = \xi\cdot X/\xi\cdot x\) when \(\xi\cdot x \not= 0\). Note
\(R(\xi) = R(t\xi)\) for any non-zero \(t\in\bm{R}\) so there is no loss
in assuming \(\xi\cdot x = 1\) when considering returns. In this case
\(R(\xi) = \xi\cdot X\) is the realized return on the portfolio. It is
common in the literature to use \emph{returns} instead of realized
returns where the return \(r\) is defined by \(R = 1 + r\Delta t\) or
\(R = \exp(r\Delta t)\) where \(\Delta t\) is the time in years or a day
count fraction of the period. Since we are considering a one period
model there is no need to drag \(\Delta t\) into consideration.

Although portfolios and prices are both vectors they are not the same. A
portfolio turns prices into a value. The function
\(\xi\mapsto \xi\cdot x\) is a \emph{linear functional} from prices to
values. Mathematically we say \(\xi\in(\bm{R}^I)^*\), the \emph{dual
space} of \(\bm{R}^I\). If \(V\) is any vector space its dual space is
\(V^* = \mathcal{L}(V,\bm{R})\) where \(\mathcal{L}(V,W)\) is the space
of \emph{linear transformations} from the vector space \(V\) to the
vector space \(W\). If we write \(\xi'\) to denote the linear functional
corresponding to \(\xi\) then \(\xi'x = \xi\cdot x\) is the linear
functional applied to \(x\). We also write the \emph{dual pairing} as
\(\langle x, \xi\rangle = \xi'x\).

Note that \(x\xi'\) is a linear transformation from \(R^I\) to \(R^I\)
defined by \((x\xi')y = x(\xi'y) = (\xi'y)x\) since \(\xi'y\in\bm{R}\)
is a scalar. Matrix multiplication is just composition of linear
operators.

\hypertarget{model-arbitrage}{%
\subsection{Model Arbitrage}\label{model-arbitrage}}

There is \emph{model arbitrage} if there exists a portfolio \(\xi\) with
\(\xi'x < 0\) and \(\xi'X(\omega) \ge0\) for all \(\omega\in\Omega\):
you make money on the initial investment and never lose money when
unwinding at the end of the period. This definition does not require a
measure on \(\Omega\).

The one-period
\protect\hyperlink{fundamental-theorem-of-asset-pricing}{Fundamental
Theorem of Asset Pricing} states there is no model arbitrage if and only
if there exists a positive measure \(\Pi\) on \(\Omega\) with
\(x = \int_\Omega X(\omega)\,d\Pi(\omega)\). We assume \(X\) is bounded,
as it is in the real world, and \(\Pi\) is a finitely additive measure.
The dual space of bounded functions on \(\Omega\) is the space of
finitely additive measures on \(\Omega\) with the dual pairing
\(\langle X,\Pi\rangle = \int_\Omega X\,d\Pi\) (Dunford and Schwartz
\protect\hyperlink{ref-DunSch63}{1963}) Chapter III.

If \(x = \int_\Omega X\,d\Pi\) for a positive measure \(\Pi\) then all
portfolios have the same expected realized return \(R = 1/\|\Pi\|\)
where \(\|\Pi\| = \int_\Omega 1\,d\Pi\) is the mass of \(\Pi\) and the
expected value is with respect to the \emph{risk-neutral} probability
measure \(Q = \Pi/\|\Pi\|\). This follows from
\(E[\xi'X] = \xi'x/\|\Pi\| = R\xi'x\) for any portfolio \(\xi\).

Note \(Q\) is not the probability of anything, it is simply a positive
measure with mass 1. The above statements are geometrical, not
probabilistic.

\hypertarget{efficient-portfolio}{%
\subsection{Efficient Portfolio}\label{efficient-portfolio}}

A portfolio \(\xi\) is \emph{efficient} if its variance
\(\operatorname{Var}(R(\xi)) \le \operatorname{Var}(R(\eta))\) for every
portfolio \(\eta\) having the same expected realized return as \(\xi\).

If \(\xi'x = 1\) then
\(\operatorname{Var}(R(\xi)) = E[(\xi'X)^2] - (E[\xi'X])^2 = E[\xi' X X'\xi] - E[\xi'X] E[X'\xi] = \xi'V\xi\),
where \(V = \operatorname{Var}(X) = E[XX'] - E[X]E[X']\). We can find
efficient portfolios using Lagrange multipliers. For a given realized
return \(\rho\) we can solve \[
    \min\frac{1}{2}\xi'V\xi - \lambda(\xi'x - 1) - \mu(\xi'E[X] - \rho)
\] for \(\xi\), \(\lambda\), and \(\mu\). The first order conditions for
an extremum are \(V\xi - \lambda x - \mu E[X] = 0\), \(\xi'x = 1\), and
\(\xi'E[X] = \rho\).

\hypertarget{riskless-portfolio}{%
\subsubsection{Riskless Portfolio}\label{riskless-portfolio}}

A portfolio \(\zeta\) is \emph{riskless} if its realized return is
constant. In this case
\(0 = \operatorname{Var}(R(\zeta)) = \zeta'V\zeta\) assuming, as we may,
\(\zeta'x = 1\). If another riskless portfolio exists with different
realized return then arbitrage exists. By removing redundant assets we
can assume there is exactly one riskless portfolio \(\zeta\) with
\(\zeta'x = 1\).

Let \(P_\parallel = \zeta\zeta'/\zeta'\zeta\). Note
\(P_\parallel\zeta = \zeta\) and \(P_\parallel\xi = 0\) if
\(\zeta'\xi = 0\) so it is the orthogonal projection onto the space
spanned by \(\zeta\). Let \(P_\perp = I - P_\parallel\) be the
projection onto its orthogonal complement,
\(\{\zeta\}^\perp = \{y\in\bm{R}^I:\zeta'y = 0\}\), so
\(V = VP_\perp + VP_\parallel\). Below we analyze the first order
conditions for an extremum on each subspace. Note \(P_\parallel\)
commutes with \(V\) so these subspaces are invariant under \(V\). Let
\(y_\parallel = P_\parallel y\) be the component of \(y\) parallel to
\(\zeta\) and \(y_\perp = P_\perp y\) be the component of \(y\)
orthogonal to \(\zeta\) for \(y\in\bm{R}^I\).

The first order condition \(V\xi = \lambda x + \mu E[X]\) implies
\(V\xi_\parallel = \lambda x_\parallel + \mu E[X]_\parallel\). Since
\(\xi_\parallel\) is a scalar multiple of \(\zeta\) we have
\(0 = \lambda + \mu R\) so \(\lambda = -\mu R\). On the orthogonal
complement \(V\xi_\perp = -\mu R x_\perp + \mu E[X]_\perp\) so
\(\xi_\perp = V^\dashv(E[X] - Rx)\) where \(V^\dashv\) is the
generalized (Moore-Penrose) inverse of \(V\). Letting
\(\alpha = \xi_\perp = V^\dashv(E[X] - Rx)\), every efficient portfolio
can be written \(\xi = \mu \alpha + \nu\zeta\). We can and do assume
\(\alpha'x = 1\) so \(1 = \mu + \nu\) and
\(\xi = \mu \alpha + (1 - \mu)\zeta\). Multiplying both sides by \(X\)
we have \(\xi'X = \mu \alpha'X + (1 - \mu)R\) hence \[
    R(\xi) - R = \mu(R(\alpha) - R).
\] This implies the classical CAPM formula by taking expected values
where \(\alpha\) is the ``market portfolio''. It also shows the Lagrange
multiplier
\(\mu = \operatorname{Cov}(R(\xi),R(\alpha))/\operatorname{Var}(R(\alpha))\)
is the classical beta.

\hypertarget{non-singular-variance}{%
\subsubsection{Non-singular Variance}\label{non-singular-variance}}

If \(V\) is invertible the
\protect\hyperlink{lagrange-multiplier-solution}{Appendix} shows
solution is \(\lambda = (C - \rho B)/D\), \(\mu = (-B + \rho A)/D\), and
\[
\xi = \frac{C - \rho B}{D}V^{-1}x + \frac{-B + \rho A}{D}V^{-1}E[X]
\] where \(A = xV^{-1}x\), \(B = x'V^{-1}E[X] = E[X']V^{-1}x\),
\(C = E[X]V^{-1}E[X]\), and \(D = AC - B^2\). The variance of the
efficient portfolio is \[
\operatorname{Var}(R(\xi)) = (C - 2B\rho + A\rho^2)/D.
\]

If \(\xi_0\) and \(\xi_1\) are any two independent efficient portfolios
then they belong to the subspace spanned by \(V^{-1}x\) and
\(V^{-1}E[X]\). Every efficient portfolio can be written
\(\xi = \beta_0\xi_0 + \beta_1\xi_1\) for some scalars \(\beta_0\) and
\(\beta_1\). Assuming \(\xi_j'x = 1\) for \(j = 0,1\) then
\(R(\xi_j) = \xi_j'X\). Assuming \(\xi'x = 1\) so \(R(\xi) = \xi'X\)
then \(\beta_0 + \beta_1 = 1\) and
\(\xi = (1 - \beta)\xi_0 + \beta\xi_1\) where \(\beta = \beta_1\).
Multiplying both sides by \(X\) we have
\(\xi'X = (1 - \beta)\xi_0'X + \beta\xi_1'X\) hence \[
    R(\xi) - R(\xi_0) = \beta(R(\xi_1) - R(\xi_0))
\] as functions on \(\Omega\) where
\(\beta = \operatorname{Cov}(R(\xi) - R(\xi_0), R(\xi_1) - R(\xi_0))/\operatorname{Var}(R(\xi_1) - R(\xi_0))\).
The classical CAPM formula follows from taking expected values of both
sides when \(\xi_1\) is the ``market portfolio'' and \(\xi_0\) is a
\protect\hyperlink{riskless-portfolio}{\emph{riskless portfolio}}.

Note that \(A\), \(B\), \(C\), and \(D\) depend only on \(x\), \(E[X]\),
and \(E[XX']\). Classical literature focuses mainly on the latter three
which may explain why prior authors overlooked our elementary but
stronger result.

\hypertarget{appendix}{%
\subsection{Appendix}\label{appendix}}

\hypertarget{lagrange-multiplier-solution}{%
\subsubsection{Lagrange Multiplier
Solution}\label{lagrange-multiplier-solution}}

Let's find the minimum value of \(\operatorname{Var}(R(\xi))\) given
\(E[R(\xi)] = \rho\). If \(\xi'x = 1\) then \(R(\xi) = \xi'E[X]\) and
\(\operatorname{Var}(R(\xi)) = \xi'V\xi\) where
\(V = E[XX'] - E[X]E[X']\).

We use Lagrange multipliers and solve \[
        \min \frac{1}{2}\xi'V\xi - \lambda(\xi'x - 1) - \mu(\xi'E[X] - \rho)
\] for \(\xi\), \(\lambda\), and \(\mu\).

The first order conditions for an extremum are \[
\begin{aligned}
        0 &= V\xi - \lambda x - \mu E[X] \\
        0 &= \xi'x - 1 \\
        0 &= \xi'E[X] - \rho \\
\end{aligned}
\] Assuming \(V\) is invertible \(\xi = V^{-1}(\lambda x + \mu E[X])\).
Note every extremum lies in the (at most) two dimensional subspace
spanned by \(V^{-1}x\) and \(V^{-1}E[X]\).

The constraints \(1 = x'\xi\) and \(\rho = E[X']\xi\) can be written \[
\begin{bmatrix}
1 \\
\rho \\
\end{bmatrix}
=
\begin{bmatrix}
\lambda x'V^{-1}x + \mu x'V^{-1}E[X] \\
\lambda E[X']V^{-1}x + \mu E[X']V^{-1}E[X] \\
\end{bmatrix}
= \begin{bmatrix}
A & B \\
B & C\\
\end{bmatrix}
\begin{bmatrix}
\lambda \\
\mu
\end{bmatrix}
\] with \(A = xV^{-1}x\), \(B = x'V^{-1}E[X] = E[X']V^{-1}x\), and
\(C = E[X]V^{-1}E[X]\). Inverting gives \[
\begin{bmatrix} \lambda \\ \mu \end{bmatrix}
= \frac{1}{D}
\begin{bmatrix}
C & -B \\
-B & A\\
\end{bmatrix}
\begin{bmatrix}
1 \\
\rho
\end{bmatrix}
=
\begin{bmatrix}
(C - \rho B)/D \\
(-B + \rho A)/D\\
\end{bmatrix}
\] where \(D = AC - B^2\). The solution is \(\lambda = (C - \rho B)/D\),
\(\mu = (-B + \rho A)/D\), and \[
    \xi = \frac{C - \rho B}{D} V^{-1}x + \frac{-B + \rho A}{D} V^{-1}E[X].
\]

A straightforward calculation shows the variance is \[
\operatorname{Var}(R(\xi)) = \xi'V\xi = (C - 2B\rho + A\rho^2)/D.
\]

\hypertarget{fundamental-theorem-of-asset-pricing}{%
\subsubsection{Fundamental Theorem of Asset
Pricing}\label{fundamental-theorem-of-asset-pricing}}

The one-period Fundamental Theorem of Asset Pricing states there is no
model arbitrage if and only if there exists a positive measure \(\Pi\)
on \(\Omega\) with \(x = \int_\Omega X(\omega)\,d\Pi(\omega)\). We
assume \(X\) is bounded, as it is in the real world, and \(\Pi\) is
finitely additive.

If such a measure exists and \(\xi\cdot X\ge0\) then
\(\xi\cdot x = \int_\Omega \xi\cdot X\,d\Pi \ge0\) so arbitrage cannot
occur. The other direction is less trivial.

\textbf{Lemma.} \emph{If \(x\in\bm{R}^n\) and \(C\) is a closed cone in
\(\bm{R}^n\) with \(x\not\in C\) then there exists \(\xi\in\bm{R}^n\)
with \(\xi\cdot x < 0\) and \(\xi\cdot y \ge0\) for \(y\in C\).}

Recall that a \emph{cone} is a subset of a vector space closed under
addition and multiplication by a positive scalar, that is,
\(C + C\subseteq C\) and \(tC\subseteq C\) for \(t > 0\). The set of
arbitrage portfolios is a cone.

\emph{Proof.} Since \(C\) is closed and convex there exists \(x^*\in C\)
with \(0 < ||x^* - x|| \le ||y - x||\) for all \(y\in C\). Let
\(\xi = x^* - x\). For any \(y\in C\) and \(t > 0\) we have
\(ty + x^*\in C\) so \(||\xi|| \le ||ty + \xi||\). Simplifying gives
\(t^2||y||^2 + 2t\xi\cdot y\ge 0\). Dividing by \(t > 0\) and letting
\(t\) decrease to 0 shows \(\xi\cdot y\ge 0\). Take \(y = x^*\) then
\(tx^* + x^*\in C\) for \(t \ge -1\). By similar reasoning, letting
\(t\) increase to 0 shows \(\xi\cdot x^*\le 0\) so \(\xi\cdot x^* = 0\).
Now \(0 < ||\xi||^2 = \xi\cdot (x^* - x) = -\xi\cdot x\) hence
\(\xi\cdot x < 0\). \(\blacksquare\)

Since the set of non-negative finitely additive measures is a closed
cone and \(X\mapsto \int_\Omega X\,d\Pi\) is positive, linear, and
continuous \(C = \{\int_\Omega X\,d\Pi : \Pi\ge 0\}\) is also a closed
cone. The contrapositive follows from the lemma.

The proof also shows how to find an arbitrage when one exists.

\hypertarget{references}{%
\subsection*{References}\label{references}}
\addcontentsline{toc}{subsection}{References}

\hypertarget{refs}{}
\begin{cslreferences}
\leavevmode\hypertarget{ref-DunSch63}{}%
Dunford, Nelson, and Jacob Schwartz. 1963. \emph{Linear Operators}. Pure
and Applied Mathematics (John Wiley \& Sons). New York: Interscience
Publishers.

\leavevmode\hypertarget{ref-Mar52}{}%
Markowitz, Harry. 1952. ``Portfolio Selection.'' \emph{The Journal of
Finance} 7 (1): 77--91.

\leavevmode\hypertarget{ref-Roy52}{}%
Roy, A. D. 1952. ``Safety First and the Holding of Assets.''
\emph{Econometrica}, no. 20: 431--49.
\end{cslreferences}

\end{document}